\documentclass[aps,prl,amsmath,amssymb,twocolumn,superscriptaddress,letterpaper]{revtex4}
\usepackage{dcolumn}
\usepackage{graphicx}
\usepackage{epstopdf}
\usepackage{verbatim}
\usepackage{amssymb}
\usepackage{amsmath}
\usepackage{subfigure,wrapfig,float}
\usepackage{color}

\begin{document}

\title{Substrate-Independent Light Confinement in Bioinspired All-Dielectric Surface Resonators}

\author{Emma C. Regan,$^{1,2\dagger}$ Yichen Shen,$^{1\star\dagger}$ Josue J. Lopez,$^{1}$ Chia Wei Hsu,$^{3}$ \\ Bo Zhen,$^{1,4}$ John D Joannopoulos,$^{1}$ and Marin Solja\v{c}i\'{c}}
\affiliation{
\normalsize{Research Laboratory of Electronics, Massachusetts Institute of Technology,}\\
\normalsize{Cambridge, MA 02139, USA}\\
\normalsize{$^{2}$Department of Physics, Wellesley College,}\\
\normalsize{Wellesley, MA 02481, USA}\\
\normalsize{$^{3}$Department of Applied Physics, Yale University,}\\
\normalsize{New Haven, CT 06520, USA}\\
\normalsize{$^{4}$Physics Department and Solid State Institute, Technion,}\\
\normalsize{Haifa 32000, Israel}\\
%\normalsize{$^{3}$Department of Mathematics, Massachusetts Institute of Technology,}\\
%\normalsize{Cambridge, MA 02139, USA}\\
%\normalsize{$^\ast$These authors contributed equally to this work.} \\
\normalsize{$^\dagger$ These authors contributed equally to the work}\\
\normalsize{$^\star$To whom correspondence should be addressed; E-mail:  ycshen@mit.edu.}
}

\begin{abstract}

Traditionally, photonic crystal slabs can support resonances that are strongly confined to the slab but also couple to external radiation. However, when a photonic crystal slab is placed on a substrate, the resonance modes become less confined, and as the index contrast between slab and substrate decreases, they eventually disappear. Using the scale structure of the \textit{Dione juno} butterfly wing as an inspiration, we present a low-index zigzag surface structure that supports resonance modes even without index contrast with the substrate. The zigzag structure supports resonances that are contained away from the substrate, which reduces the interaction between the resonance and the substrate. We experimentally verify the existence of substrate-independent resonances in the visible wavelength regime. Potential applications include substrate-independent structural color and light guiding. 

\end{abstract}

\maketitle

Interaction of light with periodic photonic structures has contributed to rapid advances in our ability to control light \cite{joannopoulos2011photonic,sakoda}, which has led to many applications, including photonic bandgap fibers \cite{Russell17012003,Knight20111998}, light management for photovoltaics \cite{lightmanagement}, and angular selective broadband reflectors \cite{Shen28032014,PhysRevB.90.125422}. Among the many photonic crystal structures, photonic crystal slabs (PhC slab) are one of the most widely used geometries due to their ease of fabrication and integration. Traditionally, a PhC slab consists of a high-index guiding layer with periodic in-plane patterns. These structures can support guided resonances that are strongly confined to the slab but also couple to external radiation \cite{PhysRevB.65.235112,Hsu2013}. The ability to channel light from the slab to the external environment has been used in optical devices such as photonic crystal surface emitting lasers \cite{phclasers} and light-emitting diodes \cite{:/content/aip/journal/apl/78/5/10.1063/1.1342048}. However, when a PhC slab is in direct contact with a substrate, the resonance modes inevitably become less confined as the index contrast between the slab and the substrate decreases. In this Letter, we present an all-dielectric surface structure that supports well-confined resonance modes without index contrast between structure and substrate; these resonances are effectively substrate-independent.

\begin{figure}[t!]
\includegraphics[width=3.25in]{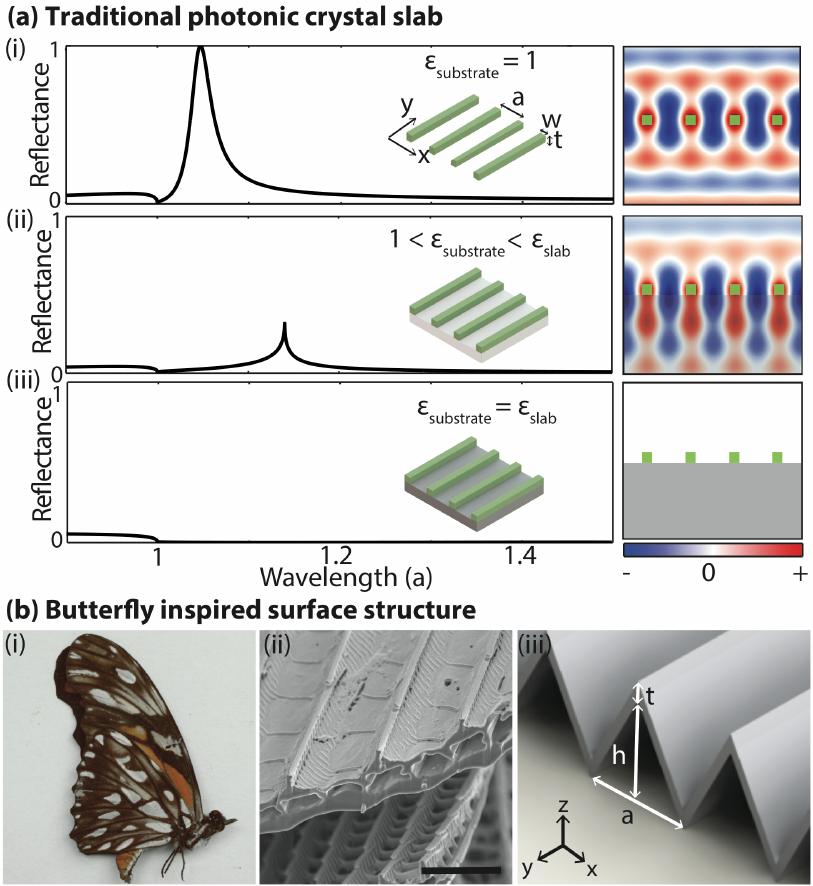}
\caption{\textbf{Illustration of resonance mode in grating and bioinspired zigzag structure.} (a) Left panel shows rigorous coupled wave analysis (RCWA) simulation of normal incidence reflectance of a grating with dielectric constant $\epsilon_{\text{slab}}=2.235$ on substrates with dielectric constant (i) $\epsilon_{\text{substrate}}=1$, (ii) $\epsilon_{\text{substrate}}=1.3$, and (iii) $\epsilon_{\text{substrate}}=2.235$ \cite{Liu20122233,Gaylord:86}. In each case, the grating is identical, with periodicity $a$, thickness  $t=0.25a$, and width $w=0.25a$, and the incident light polarized in the y-direction. In the right panel (i) and (ii) show finite difference time domain (FDTD) mode profile ($E_y$) \cite{FDTD,OskooiRo10}. (iii) shows no resonance mode is supported due to strong leakage to the substrate. (b) (i) Photograph of the Dione juno butterfly's wings. (ii) SEM image of the cross section (focus ion beam milled) of a cut silver scale and wing substrate, showing that the scale body forms a periodic pattern on a substrate with air gaps. Scale bar is $2~\text{$\mu$m}$ (iii) Schematic illustration of the abstracted zigzag structure characterized by periodicity $a$, height $h$, and thickness $t$.}
\label{fig:fig1}
\end{figure}

When a PhC slab is on a substrate that has a higher refractive index than the environment, the resonance modes inside the slab typically couple to the substrate, reducing their lifetimes. To illustrate this phenomenon, consider a 1D dielectric grating under vacuum [Fig. \ref{fig:fig1}(a)]. With no substrate, the grating supports a first-order resonance mode whose energy is concentrated within the slab. Reflection spectra provide experimental evidence of these resonance modes. When the PhC slab is illuminated from above, the radiated light from the resonance mode interferes with the directly reflected light, forming a sharp optical Fano resonance reflection peak \cite{PhysRevB.65.235112,RevModPhys.82.2257,doi:10.1021/ph500400w}. As the dielectric constant of the substrate increases, two factors increase the radiation rate of the resonance \cite{PhysRevB.85.235145}. First, the number of radiation channels increases. Second, the impedance mismatch between the slab and the substrate decreases, making the radiation stronger for each channel. As a result, the lifetime of the resonance mode is typically reduced. The mode leakage increases with the dielectric constant of the substrate until no resonance modes are supported once the dielectric constant of the substrate equals that of the slab. At this point, the resonant properties of the slab are no longer observable, as shown in Fig. \ref{fig:fig1}(a(iii)) \cite{subnote}. Furthermore, the resonances shift spectrally when a substrate is introduced due to the modified effective dielectric constant.

To reduce the substrate leakage, metals or high-index dielectric materials are typically used. While plasmonics promise control of light on the subwavelength scale, losses in the optical and infrared regions often significantly limit mode lifetimes and device performance \cite{plasmonic_loss}. A small number of high-index dielectrics, such as silicon, are used to efficiently guide light in the telecommunication wavelength regime, but these materials are lossy at visible wavelengths. Materials that have low absorption in the visible spectrum, such as SiN and $\text{TiO}_\text{2}$, have comparatively low dielectric constants. While low-index waveguides have been demonstrated using photonic band gaps or index contrast \cite{Xu:04,Abouraddy2007,Litchinitser:02,RevModPhys.82.2257}, the trade-off between substrate leakage and optical loss seems inevitable for surface resonators in the visible spectrum. Overcoming this constraint is important for many applications, such as on-chip biosensing, which relies on low optical absorption in water \cite{biosensing}. In light of the limitations of traditional PhC slabs and their alternatives, we introduce a new class of bioinspired, low-index surface resonators that uses periodic air gaps to support resonance modes without index contrast between the structure and the substrate.

Photonic crystal structures in butterfly wings use low-index biological materials to produce dramatic colors \cite{photonics_bio} and have been fabricated for optical devices \cite{ANIE:ANIE201103505,ADFM:ADFM201102948}. We study the silver scales on the \textit{Dione juno} butterfly as an inspiration for a surface resonator with low index contrast. A scanning electron micrograph of the cross section of the silver scales and wing substrate shows two ridged cellular lamina with periodic connections [Fig. \ref{fig:fig1}(b(ii))]. If the periodic structure interacts with light, the air gaps between the two laminae may reduce the radiation of a top-layer resonance into the underlying lamina structure, despite the low index contrast between the two. 

Inspired by the scales, we abstract the microscopic structure to a zigzag surface structure characterized by its height (h), thickness (t), and periodicity (a), whose structure mimics the periodic air gaps in the scales [Fig. \ref{fig:fig1}(b(iii))]. This structure can be fabricated using direct laser writing or with controlled buckling for large scale fabrication \cite{nanoscribe,doi:10.1021/am4054207}. Moreover, the zigzag structure may be suitable for dynamic tuning if fabricated on a stretchable substrate, which merits further investigation.

\begin{figure*}[hbtp!]
\centering
\includegraphics[width=7in]{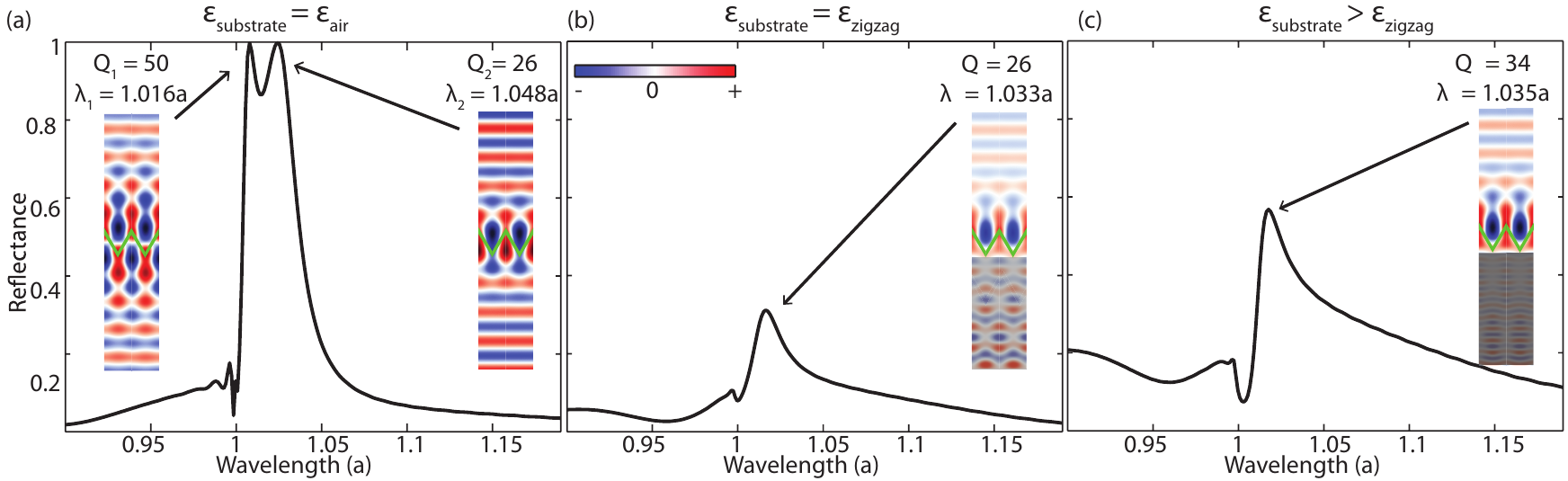}
\caption{\textbf{Numerical simulation of the zigzag structure.} FDTD simulation of the reflectance of acrylic $\epsilon_{\text{zigzag}}=2.235$ zigzag structure with periodicity $a$, height $h=0.88a$, thickness $t=0.22a$, and dielectric constants (a) $\epsilon_{\text{substrate}}=\epsilon_{\text{air}}=1$, (b) $\epsilon_{\text{substrate}}=\epsilon_{\text{zigzag}}=2.235$, (c) $\epsilon_{\text{substrate}}=6>\epsilon_{\text{zigzag}}$. The incident light is polarized in the y-direction. Inset in each figure: FDTD mode profiles ($E_y$) and Q-factors of the resonance modes at wavelength equal to the first-order Fano resonance peaks indicated.}
\label{fig:fig2}
\end{figure*}

We study the resonant properties of the zigzag structure with the finite difference time domain (FDTD) method \cite{FDTD}, using a freely-available software package \cite{OskooiRo10}. We use periodic boundary conditions in the plane of the periodicity and perfectly matched layer (PML) boundary conditions above and below the structure. A low-index acrylic ($\epsilon=2.235$) zigzag structure was chosen that produces a strong reflection peak on high-index substrates. The structure has periodicity $a$, height $h=0.88a$, and thickness $t=0.22a$. Reflection spectra were calculated for the zigzag structure on three substrates: vacuum ($\epsilon=1$), acrylic ($\epsilon=2.235$), and high-index ($\epsilon=6$) [Fig. \ref{fig:fig2}]. Using a low-storage filter diagonalization method, the spectral location and quality factors ($\text{Q-factor}\approx\omega_{0}/\gamma$, where $\omega_{0}$ and $\gamma$ are the frequency and width of the resonance) of the supported resonance modes were determined for each zigzag-substrate combination \cite{harminv}. The wavelengths of the resonances agree well with the resonant reflection peaks, indicating that the peaks are caused by optical Fano resonance. The mode profiles of the resonances are shown in the inset of Fig. \ref{fig:fig2}. 

The numerical results show that for $\epsilon_{\text{substrate}}=1$, the acrylic zigzag structure supports two resonance modes at wavelengths $\lambda_1=1.016a$ and $\lambda_2=1.048a$. For $\epsilon_{\text{substrate}}>1$, one resonance remains near this wavelength range, whose spectral location is approximately constant for different substrate dielectric constants: on the acrylic substrate, $\lambda=1.033a$, and on the high-index substrate, $\lambda=1.035a$. This resonance remains for substrates with $\epsilon_{\text{substrate}}\geq\epsilon_{\text{slab}}$, which is not possible in traditional PhC slabs. The other resonance leaks into the substrate. Additionally, in traditional PhC slabs, the resonant wavelength undergoes a large shift as the substrate index increases [Fig. \ref{fig:fig1}(a)]. The constant mode location in the zigzag structure suggests that the resonance mode is weakly interacting with the substrate. This is confirmed by the mode profiles inside the zigzag structure, which show that the resonance mode is localized away from the substrate and is concentrated around the horizontal plane containing the top points of the zigzag due to the large volume of high-index material contained away from the substrate. Therefore, the coupling is reduced and the effective index that the mode experiences is roughly constant [Fig. \ref{fig:fig2}(b-c)]. Unlike in traditional photonic crystal slabs, the Q-factor increases slightly for higher-index substrates because the reflection at the air/substrate interface increases. The robustness of the Q-factor and the resonance peak location suggest that these resonances may be useful for devices that work on a variety of substrates.

The substrate-independent resonance property is not unique to the zigzag geometry. A different abstraction of the wing structure, a lamina with periodic connections to the substrate, also shows a sustained resonance on high-index substrates [Fig. S1]. Again, this structure supports a resonance mode away from the substrate with low coupling because the slab material is concentrated away from the substrate \cite{10.1038/ncomms6718}. To better understand this physical system, we note that this structure and the zigzag both contain periodic air gaps between the mode and the substrate. Therefore, the surface resonator with self-contained air gaps is similar to a suspended traditional PhC slab, where the distance between the slab and the substrate determines the coupling [Fig. S2].

\begin{figure*}[t!]
\centering
\includegraphics[width=7in]{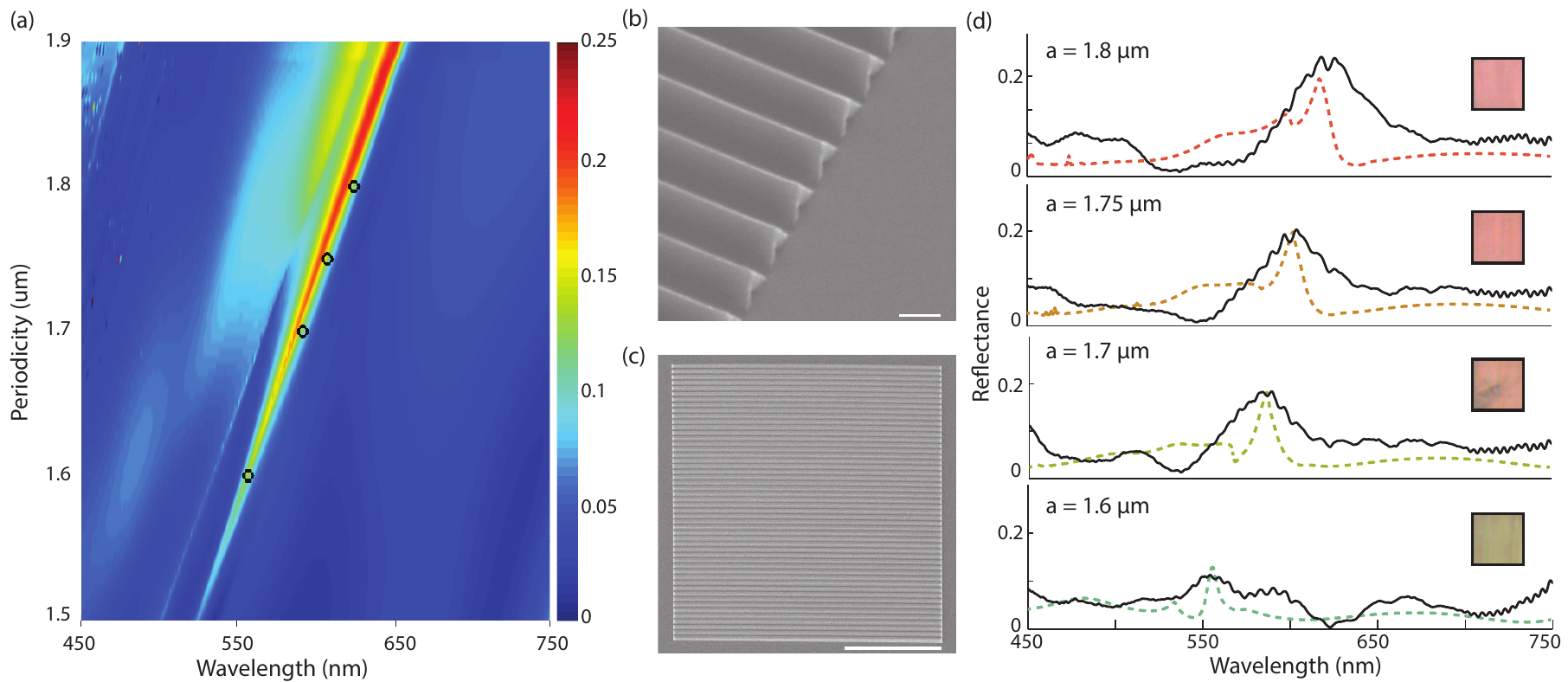}
\caption{\textbf{Experimental design and measurement.} (a) RCWA simulation of the reflectivity into the normal direction of a zigzag structure with different periodicities $a$ and fixed thickness $t=0.3~\text{$\mu$m}$ and height $h=1.6~\text{$\mu$m}$.Incident light is unpolarized. The black circles indicate the spectral location of third-order resonance modes inside the zigzag structure at each fabricated periodicity, determined by a low-storage filter diagonalization method. (b-c) SEM image of the cross section and the top view of a fabricated zigzag structure on a fused silica substrate. Scale bar is $2~\text{$\mu$m}$ in (b) and $50 ~\text{$\mu$m}$ in (c). (d) RCWA simulation (dashed line) and experimental measurement (solid line) of the reflectivity into the normal direction of the fabricated zigzag structures with different periodicities. Incident light is unpolarized. The colors of the dashed lines correspond to the perceived colors. The images inside the black boxes are the microscope photographs of the measured samples ($50~\text{$\mu$m}\times50~\text{$\mu$m}$).}
\label{fig:fig3}
\end{figure*}

To verify the existence of such resonances, we experimentally measured the reflection spectrum of zigzag structures on a substrate with little index contrast and observed the optical Fano resonance peaks caused by the predicted resonance modes. The experiment was conducted in the visible spectrum, where alternative methods (plasmonics and high-index dielectrics) typically perform poorly because of losses. A zigzag structure with $\epsilon=2.31$ was optimized to support one set of third-order resonances in the visible spectrum when on a fused silica ($\epsilon=2.18$) substrate [Fig. \ref{fig:fig3}(a)]; third-order resonances were necessary to study reflection peaks in the visible spectrum because the spatial resolution of the fabrication process is $300~\text{nm}$. Zigzag structures with thickness $t=0.3~\text{$\mu$m}$, height $h=1.6~\text{$\mu$m}$ and periods $1.6$, $1.7$, $1.75$, and $1.8~\text{$\mu$m}$ were fabricated using direct laser writing. A fused silica substrate ($\epsilon=2.18$) was coated with a liquid IP-L 780 photoresist, which was exposed to a $780~\text{nm}$ excitation laser with a pulse width of $100~\text{fs}$ and a repetition rate of $80~\text{MHz}$ \cite{nanoscribe}. The sample was then developed with 1-methoxy-2-propanol-acetate. The photoresist, which forms the zigzag structure, has a dielectric constant $\epsilon=2.31$ after laser exposure. Samples of $50~\text{$\mu$m}\times50~\text{$\mu$m}$ were fabricated, which includes about $30$ periods. Scanning electron micrographs show that zigzag patterns were achieved [\ref{fig:fig3}(b-c)]. Alternatively, large-area fabrication of zigzag surface structures may be possible using controlled buckling, which has produced PDMS zigzag micropatterns with wavelength of $\approx50~\text{$\mu$m}$ \cite{doi:10.1021/am4054207}. If the micropattern is attached to a substrate, this fabrication technique could be used to create large-scale spectrum filters. 

Reflection measurements near the normal incidence of the zigzag structures were performed using a microspectromer (CRAIC QDI 2010) with a collection range of $\pm 2^{\circ}$ [Fig. S3]. The imaging spot size was approximately $40~\text{$\mu$m}\times40~\text{$\mu$m}$. The color of each structure is easily visible through the microscope [\ref{fig:fig3}(d)]. The measured reflectance spectra of samples fabricated with different periodicities are shown in Fig. \ref{fig:fig3}(d), along with rigorous coupled wave analysis (RCWA) simulations of normal incidence reflection \cite{Liu20122233}. The experimental reflection peaks are broadened due to disorder-induced scattering and side leakage due to the finite size of the structure \cite{PhysRevLett.94.033903,PhysRevB.72.161318}. Additionally, the experimental peaks are higher than predicted due to the small range of collection angles around the normal direction. The dimensions of the zigzag structure are at the limit of the resolution obtainable from the direct laser writing process. Therefore, we expect some fabrication inaccuracies, which are more significant for smaller structures. Despite this limitation, we observe reflection peaks that agree closely with the numerical reflection spectra. The resonance wavelength increases with the period of the zigzag structures, as expected. The appearance of the reflection peaks experimentally validates the existence of well-confined substrate-independent resonances with low index contrast.

The substrate-independent properties are useful for many applications. One application is a single-material, low-index structural color mechanism. Structural color has previously been produced using light interaction with bulk materials via mechanisms that include scattering \cite{ADMA:ADMA200903693}, photonic crystals \cite{ADEM:ADEM201300089}, and multilayer interference \cite{Kolle:10}. However, these techniques require thick bulk material to generate sufficient reflection. Plasmonic resonances have also been used to produce color in nanoscopically structured metal surfaces \cite{doi:10.1021/nl501460x}. These techniques promise low thickness and a wide range of color, but the reflection contrast is low because metals are generally lossy in the visible spectrum. A Fano resonance structural color mechanism has been studied in silicon PhC slabs \cite{doi:10.1021/ph500400w}, but was not previously possible with lossless low-index materials. This method creates strong reflection peaks for $\lambda>600~\text{nm}$, but increased absorption in silicon at smaller wavelengths reduced the spectral range of color. In contrast, the resonance in a dielectric zigzag structure persists on a substrate of the same material and produces broad Fano resonance peaks that are suitable for color in the full visible spectrum. The color can be selected via the periodicity of the zigzag, as has been shown in our experiment. Similar to previous silicon-based Fano resonance structural color results, the zigzag structure presented here could be further optimized for even stronger reflection peaks \cite{doi:10.1021/ph500400w}. 

Substrate-independent resonances may also find applications in photonic integrated circuits (PICs), which perform a variety of optical functions on a single chip and have shown promise for many applications \cite{C0AN00449A,5340692}. While silicon dominates the semiconductor industry, PICs have been fabricated using a variety of host material systems that provide unique functionality based on their material properties \cite{5069750}. A low-loss waveguide that operates using self-contained periodic air gaps, rather than relying on high index contrast between the waveguide and the surrounding medium, is an interesting alternative to the material-specific waveguide designs used today. Further, while resonance modes in plasmonic structures have limited quality factors due to absorptive losses, \cite{Hsu2013,ADFM:ADFM201000135,Yanik19072011} the resonance mode lifetimes for substrate-independent waveguides depend only on substrate leakage and can be optimized geometrically. Optimization of slab with periodic air gaps that supports high-Q guided modes would further contribute to a new class of all-dielectric structures that have substrate-independent properties and work at visible wavelengths. As PICs become increasingly important for high-speed communication and efficient sensing, new methods for designing passive optical components may provide a way to overcome material constraints.

In this Letter, we introduce a new class of low-index bioinspired surface structures that support resonance modes on high-index substrates and without index contrast, which was not previously possible. We experimentally verify the existence of such resonances in the visible wavelength regime, and we identify potential applications for structural colors and light guiding. 

\section*{Funding Information}
This work was partially supported by the Army Research Office through the ISN under contract nos. W911NF-13-D0001. The fabrication part of the effort was supported by the MIT S3TEC Energy Research Frontier Center of the Department of Energy under grant no. DE-SC0001299. B.Z. was partially supported by the United States-Israel Binational Science Foundation (BSF) under award no. 2013508. J.J.L. is supported by the NSF Graduate Research Fellowship under award no. 1122374 and an MIT Presidential Lemelson Fellowship. Part of the work was performed at the Harvard Center for Nanoscale Systems (CNS), a member of the National Nanotechnology Coordinated Infrastructure Network (NNCI), supported by the National Science Foundation under NSF award no. 1541959.

\section*{Acknowledgements} Many thanks to Conrad Stansbury for helpful revisions. 

\bibliography{regan_manuscript}

\section*{Supplementary}

\renewcommand{\thefigure}{S\arabic{figure}}
\setcounter{figure}{0}

\begin{figure*}[htbp]
\centering
\fbox{\includegraphics[width=\linewidth]{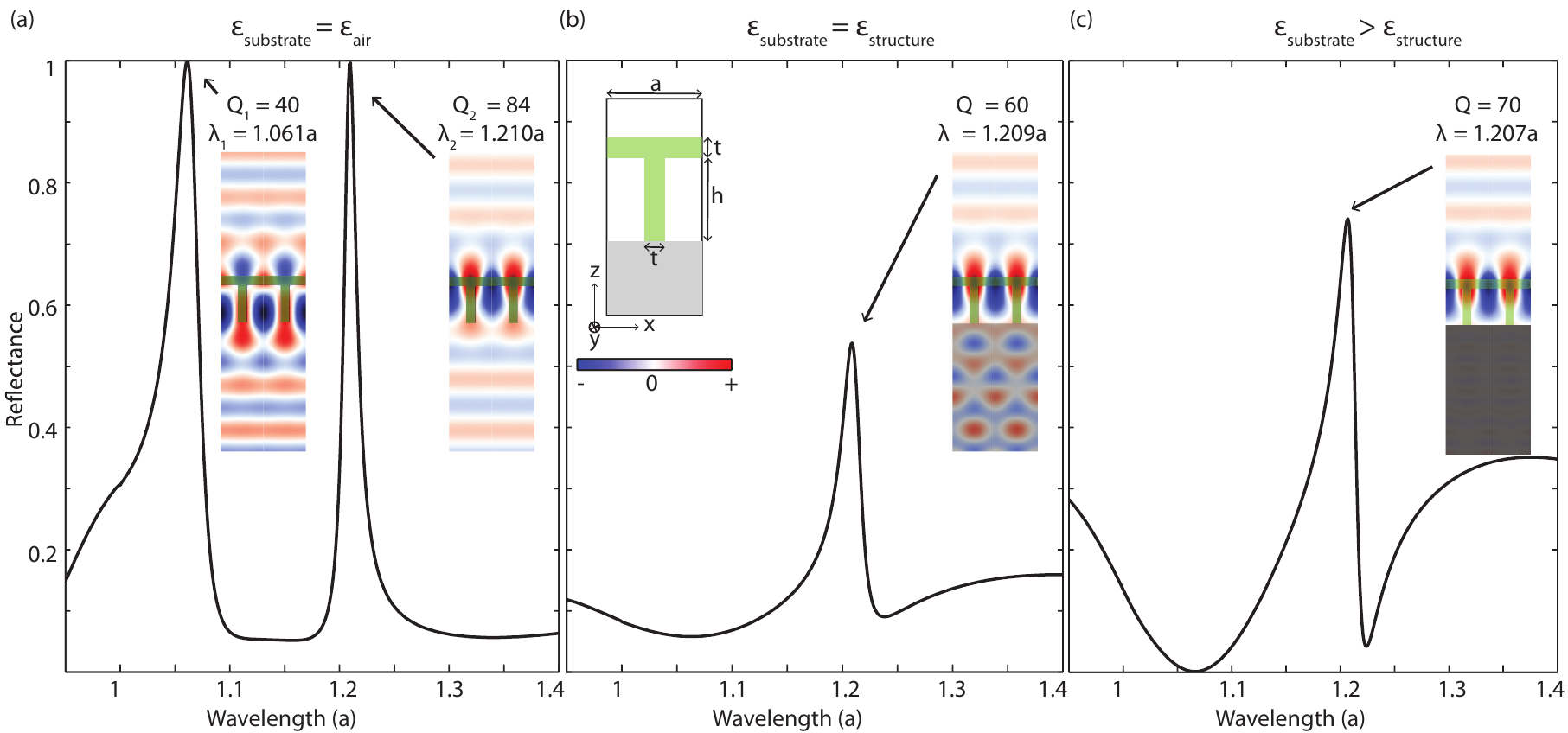}}
\caption{\textbf{Numerical Simulation of T-bar Structure.} FDTD simulation of the reflectance of an acrylic $\epsilon=2.235$ T-bar structure (see inset of (b)) with periodicity $a$, height $h=0.88a$, thickness $t=0.22a$ on substrate with dielectric constants (a) $\epsilon_{\text{substrate}}=\epsilon_{\text{air}}=1$, (b) $\epsilon_{\text{substrate}}=\epsilon_{\text{t-bar}}=2.235$, (c) $\epsilon_{\text{substrate}}=6>\epsilon_{\text{t-bar}}$. The incident light is polarized in the y-direction. Inset in each figure: FDTD mode profiles ($E_y$) and Q-factors of the resonance modes at wavelengths equal to the first-order Fano resonance peaks indicated.}
\label{fig:sup1}
\end{figure*}

\begin{figure*}[htbp]
\fbox{\includegraphics[width=\linewidth]{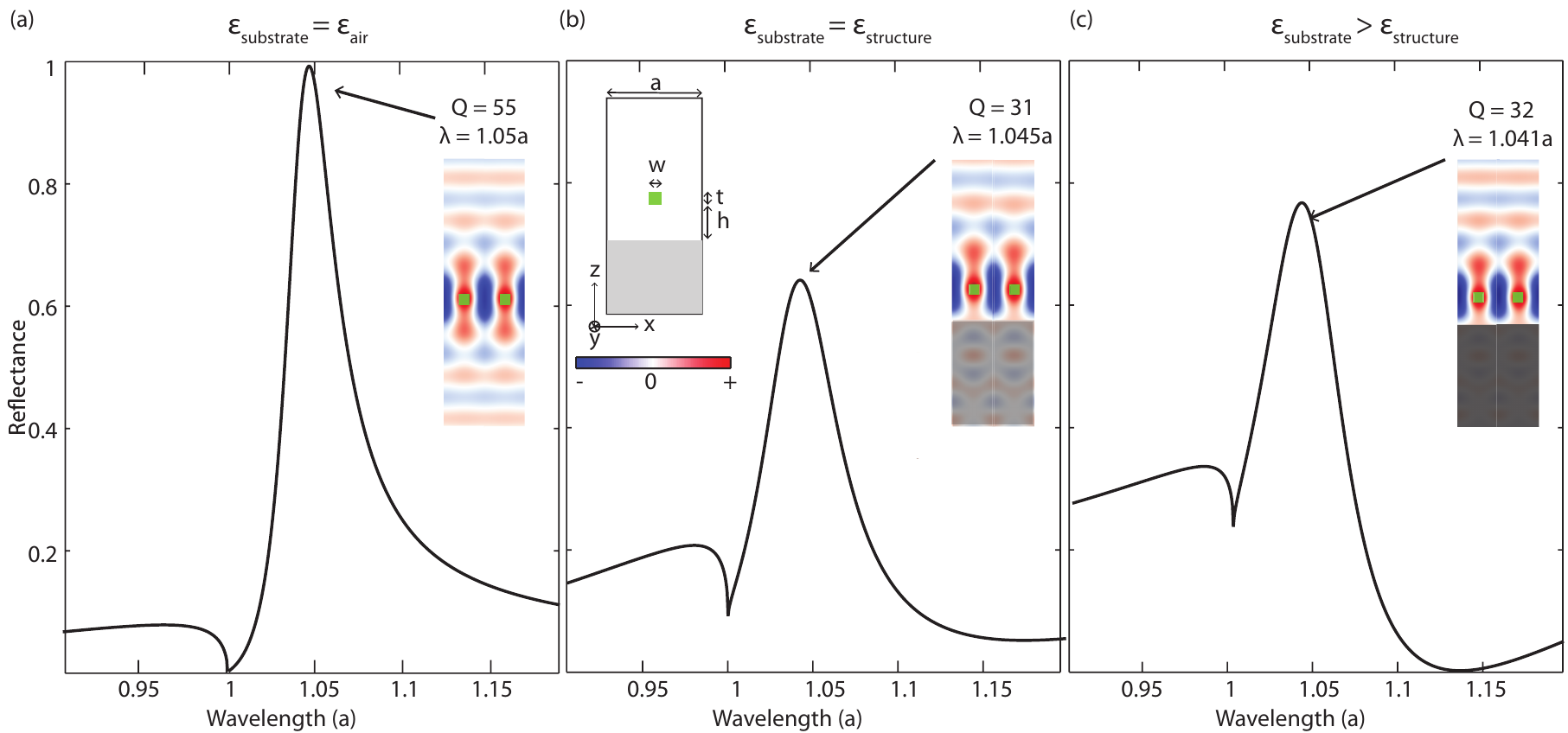}}
\caption{\textbf{Numerical Simulation of Suspended Grating.} RCWA simulation of the reflectance of an acrylic $\epsilon=2.235$ grating (see inset of (b)) with periodicity $a$, width $w=0.22a$, thickness $t=0.22a$ suspended at $h=0.56a$ above a substrate with dielectric constants (a) $\epsilon_{\text{substrate}}=\epsilon_{\text{air}}=1$, (b) $\epsilon_{\text{substrate}}=\epsilon_{\text{slab}}=2.235$, (c) $\epsilon_{\text{substrate}}=6>\epsilon_{\text{slab}}$. The incident light is polarized in the y-direction. Inset in each figure: FDTD mode profiles ($E_y$) and Q-factors of the resonance modes at wavelengths equal to the first order Fano-resonance peaks indicated.}
\label{fig:sup2}
\end{figure*}

\begin{figure*}[htbp]
\centering
\includegraphics[width=\linewidth]{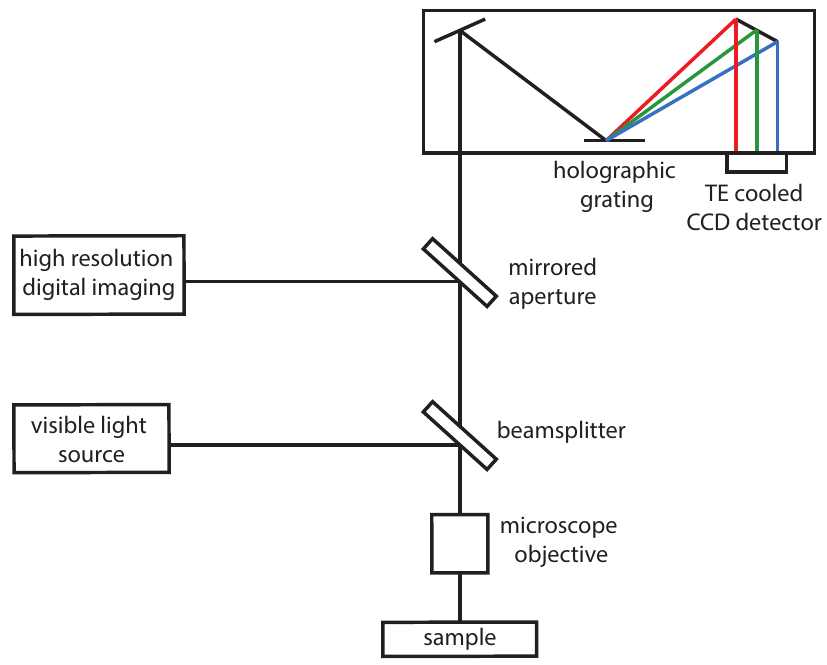}
\caption{\textbf{Microspectrometer Setup for Reflectance Measurement.} Schematic of CRAIC QDI 2010 microspectrometer used to excite zigzag sample at normal incidence and collect reflected light at normal $\pm2^{\circ}$.}
\label{fig:sup3}
\end{figure*}

\end{document}